\documentclass[fleqn,11pt]{wlscirep}
\title{Funding shapes the anatomy of scientific research}

\author[1,*]{Athen Ma}
\author[1]{Ra\'ul J. Mondrag\'on}
\author[2]{Vito Latora}
\affil[1]{Queen Mary University of London, School of Electronic Engineering and Computer Science, London E1 4NS United Kingdom}
\affil[2]{Queen Mary University of London, School of Mathematical Sciences, London E1 4NS United Kingdom}

\affil[*]{athen.ma@qmul.ac.uk}

\def\degree{{k}}
\def\linksRC{k^+}

\def\degree{k}

\newcommand{\beginsupplement}{%
    	\setcounter{table}{0}
    	\renewcommand{\thetable}{S\arabic{table}}%
    	\setcounter{figure}{0}
    	\renewcommand{\thefigure}{S\arabic{figure}}%
 	}

\usepackage{color}
\usepackage{longtable}

\keywords{Keyword1, Keyword2, Keyword3}

\begin{abstract}
Research projects are primarily collaborative in nature through
internal and external partnerships, but what
role does funding play in their formation? Here, we examined over 43,000 funded projects in the past three decades, enabling us to characterise changes in the funding landscape and their impacts on the underlying collaboration patterns. We observed rising inequality in the distribution of funding and its effect was most noticeable at the institutional level in which the leading universities diversified their collaborations and increasingly became the knowledge brokers. Furthermore, these universities formed a cohesive core through their close ties, and such reliance appeared to be a key for their research success, with the elites in the core over-attracting resources but in turn rewarding in both research breadth and depth. Our results reveal how collaboration networks undergo previously unknown adaptive organisation in response to external driving forces, which can have far-reaching implications for future policy.
\end{abstract}
\begin{document}

\flushbottom
\maketitle
%
%
\thispagestyle{empty}


\section*{Introduction}

Higher education institutions are nationally assessed in a periodic manner across the globe, examples include the Research Excellence Framework~\cite{REF} in the United Kingdom, Excellenzinitiative~\cite{Excellen} in Germany and Star Metrics~\cite{StarMetrics} in the United States, and tremendous effort have been put in place in maximising research output as assessment outcomes often have a direct financial impact on revenue~\cite{geuna2006university,leru2012}. Bibliometrics are commonly used for this kind of performance evaluations~\cite{barabasi2012publishing,wang2013quantifying,deville2014career}, and the volume of grant income is generally seen as a good indicator of performance. While many studies have examined the collaboration patterns originated from publication information~\cite{moody2004structure,newman2004best,newman2004coauthorship,wagner2005network,glanzel2005analysing,guimera2005team,sonnenwald2007scientific}; little is known about the characteristics of project collaborations supported by research funding, which is undoubtedly a type of research output in its own right, but also the origin of other research outputs. 

The availability of funding is often subject to direct and indirect constraints arisen from internal research strategy and different levels of policy set out by the funding bodies and ultimately the government, manifesting into different emphasises on both the research area and mode of collaboration, and this potentially influences the way we form a project team.  We have already seen examples of adaptive changes in our collaboration practises; for instance, research in the science and engineering sector is said to be increasingly inter-organisational~\cite{jones2008multi}. In addition, there are different theories on the factors that may affect the establishment of a collaboration and how well a research team operates. Elite universities were recognised as catalysts for facilitating large scale multi-partner research collaborations~\cite{jones2008multi}, and multidisciplinary collaborations were found to have higher potential to foster research outcomes~\cite{cummings2005collaborative}. As a result, the setup of a project consortium for a grant application might require considerable strategic planning, as who and how we collaborate can potentially impact the outcome of a bid, and we are yet to fully understand the mechanistic for success.

In order to shed light into the relations between funding landscapes and scientific collaborations, here we examined over 43,000 collaborative projects funded between 1985 and 2013 by the Engineering and Physical Sciences Research Council (EPSRC), the government body in the United Kingdom that provides funding to universities to undertake research and postgraduate degrees in engineering and the physical sciences, including mathematics, chemistry, materials science, energy, information and communications technology, and innovative manufacturing. For each year, we constructed two different types of collaboration networks in which the nodes are investigators and their affiliations respectively, and an edge represents a funded project partnership between two nodes. We applied a novel network-based approach to analyse the local and global interlinkage in these networks; the former was performed by calculating the effective size~\cite{ronald1992structural,latora2013social} of individual nodes, which gauges the connectivity in the neighbourhood of a node. As for the global level, we calculated the rich-club coefficient~\cite{zhou2004rich, colizza2006detecting} of the network and characterised the members of such core structure using a novel profile technique~\cite{ma2014rich}. In addition, we explored how these patterns evolved over time with the availability of funding and how they correlated with research performance~\cite{citation,hirsch2005index, hirsch2007does}. Our results allow us to gain an insight into how changes in the funding landscape shaped the way we form research partnerships, providing a case study that is highly reflective of other countries in the European Union and possibly other developed countries world wide.

\section*{Results}

\subsection*{Changes in the funding landscape}
\begin{figure}[!htb]
\centering
\includegraphics[width=10cm]{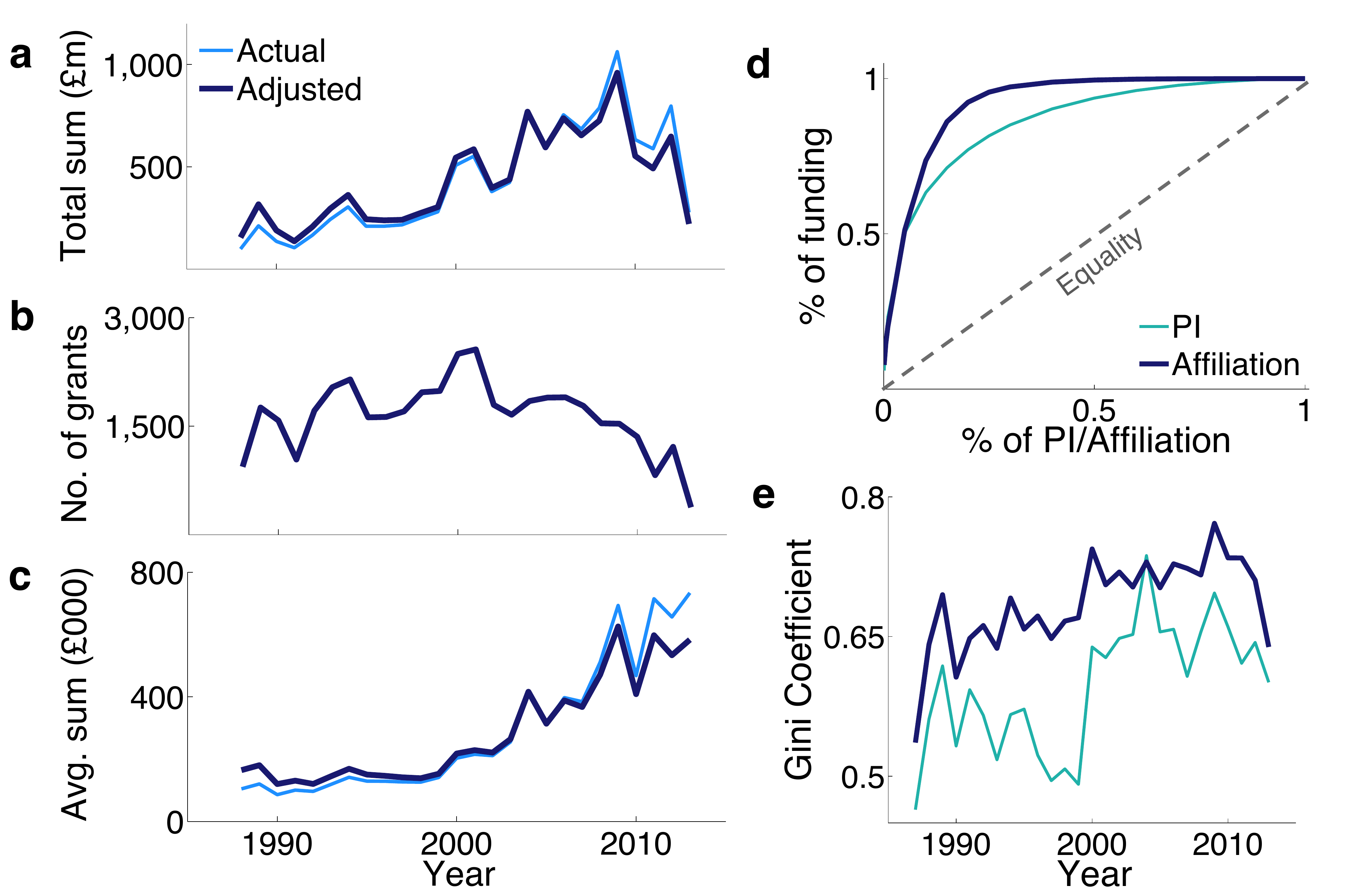}
\caption{\textbf{The funding landscape over the last three decades.}
(a). The total amount
  of funding awarded by EPSRC showed a general tendency to increase,
  with a peak in 2009, and then a decrease. \emph{Actual} refers to
  the actual value on record while \emph{adjusted} refers to the value
  after the adjustment made with reference to the Consumer Prices Index. (b). The total number of grants awarded by EPSRC
  peaked in 2001. (c). The average amount of funding per
  grant continued to rise over time. (d). Overall distribution
  of funding among PIs and affiliations. Individual awardees (PIs or
  affiliations) were sorted in descending order of their total
  funding, and the percentage of funding was plotted against the
  corresponding percentage of awardees. The dotted line denotes where true equality lies and both cases showed strong inequality. (e). Distribution of
  funding consistently showed a high degree of inequality over time,
  in both cases of PIs and affiliations.}
\label{fig:grant}
\end{figure}

During the period of study we found that the overall funding increased steadily over time,
until it peaked in 2009 (Fig.~\ref{fig:grant}a). The number of grants fluctuated over time with a general decline after 2001 (Fig.~\ref{fig:grant}b) and, in essence, there was an obvious trend of fewer grants but of much larger monetary values (Fig.~\ref{fig:grant}c). This coincided with the emergence of larger research teams (see Supplementary Fig. S1 online) and the timing of this tied in with the EPSRC's initiative on developing larger specialist units in the UK, such as establishing Doctoral Training Centres in selected universities~\cite{EPSRC2010}. Investigators associated with a grant were classified into Principal Investigators (PIs),
Co-Investigators (CIs) and  Other Investigators (OIs). There were a total of 13,275 PIs from 201 different affiliations (out of a total of 1834), and the average number of grants per PI was 3.25. About half of the grants were associated with one or more CIs and/or OIs, and there was a noticeable rise in the average numbers of investigators since 2000, which is in line with the observed increase in the typical number of collaborators in scientific publications~\cite{grossman1995portion,newman2004best}. Though, we did not observe the same degree of widening participation in affiliations as the average number of affiliations associated to a grant only marginally increased (see Supplementary Fig. S1 online). 

The way in which funding has been awarded was highly skewed, as we found that in both cases of PIs and affiliations about half of the available funding was awarded to the top 8\% (Fig.~\ref{fig:grant}d); though the distribution of funding among the remaining PIs showed a much greater homogeneity. On the contrary, we observed over 90\% of all the funding was awarded just to the top 20\% of the affiliations, suggesting a high level of focussed funding in selected places. Furthermore, we referred to the Gini
coefficient~\cite{RePEc:oxp:obooks:9780198281931} in order to measure how the  distribution of funding over a population deviates from a perfectly uniform distribution, with values $G=0$ and $G=1$ denoting maximum equality and inequality respectively. Overall, we observed substantial inequality in both cases of  PIs and affiliations of PIs. By examining the Gini coefficient over time, we found that the level of inequality intensified as the coefficient became closer to 1 (Fig. \ref{fig:grant}e), and this was particularly the case among the affiliations.

\subsection*{Network brokerage}

\begin{figure}[!h]
\centering
\includegraphics[width=17.5cm]{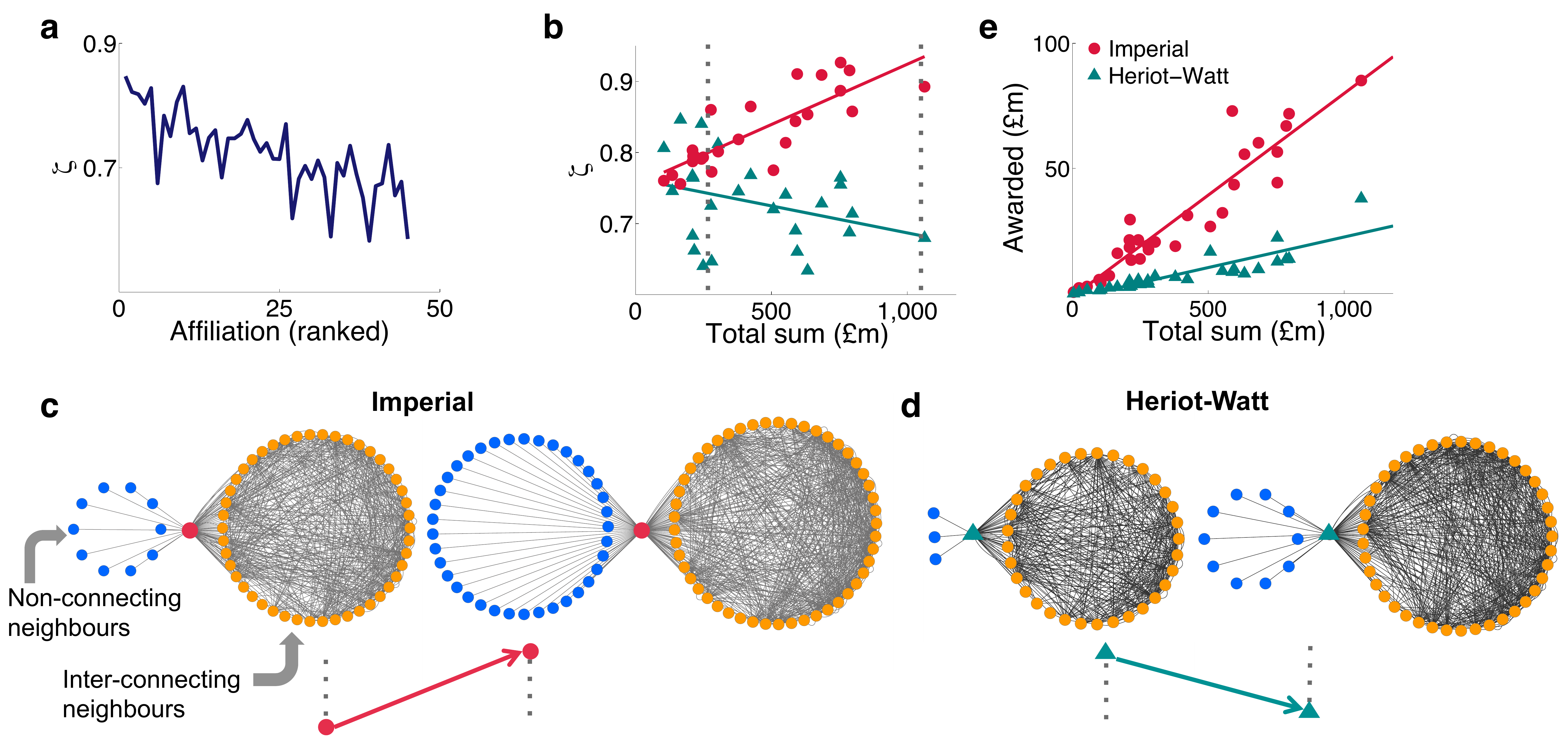}
\caption{\textbf{The brokers of the affiliation network.} (a).
The level of brokerage, as measured by the yearly average effective size $\zeta$, is
reported for the 50 top-funded affiliations ranked in descending order of their total funding. (b). The level of brokerage of Imperial and Heriot-Watt exhibited opposites trends with the rise of funding. (c). Connectivity between Imperial College and its partners when the total sum was \pounds250 million and over \pounds1000 million as indicated. Both non-connecting and inter-connecting neighbours of Imperial College are shown.  (d). Same as (c) but for Heriot-Watt University and its partners. (e). Funding obtained by Imperial and Heriot-Watt continued to rise linearly with the total funding available.}
\label{fig:richcore}
\end{figure}

We examined the two constructed collaboration networks in which nodes are investigators and affiliations, respectively, and an edge represents a funded partnership between two nodes, and found that network properties evolved over time depending on the total available funding (see Supplementary Fig. S2 and Fig. S3 online). The investigator network appeared to be sparsely connected and consisted of a large number of disjoint parts, which means that collaborations were largely localised to small groups of investigators. In contrast, the affiliation network was found to be well connected with a definitive giant component. To further examine this noticeable level of interconnectedness in the affiliation network, we examined the extent of local cohesion between affiliations and their partners in the form of brokerage~\cite{ronald1992structural,latora2013social}; a broker is said to occupy an advantageous location in the network for detecting and developing opportunities through their connections to non-overlapping clusters. This was performed by computing the
\emph{normalised effective size} $\zeta_i$ of the neighbourhood of each affiliation $i$ in the collaboration network. Such a quantity ranges from 0 to 1. It takes its smallest value when
$i$ is part of a clique (i.e. a fully cohesive structure), while it is equal to 1 when $i$ is the centre of a star, and there is no link between any of its partners. Generally, the larger the value of $\zeta_i$, the less connected the neighbourhood of $i$ is, and consequently, the higher the \emph{brokerage opportunities} of affiliation $i$. Overall, we found that the effective size of affiliations increases with the total funding (Fig.~\ref{fig:richcore}a), with the top-funded universities occupying brokerage
positions between otherwise disconnected affiliations, potentially playing a key role in developing new access to information and opportunities.

In order to gain a better insight into the factors that may contribute towards success in terms of awarded grants, we examined how the brokerage behaviour of the leading affiliations has changed with the total funding over the last three decades. We found that the rise in the volume of the total funding,
which coincided with the emergence of investment on focussed research, has resulted in further centralisation of the local networks of the top 10 funded affiliations, as  
reflected by the increase of $\zeta$ with funding. This is clearly shown in the case of Imperial College (Fig.~\ref{fig:richcore}b), as it increasingly acted as information brokers between otherwise unconnected neighbours (Fig.~\ref{fig:richcore}c); most likely thanks to their ability in facilitating new research partnerships~\cite{jones2009burden}.
On the contrary, the lesser funded affiliations were found to exhibit the opposite trend, as shown in the case of Heriot-Watt University (Fig.~\ref{fig:richcore}b); as their $\zeta$ decreased with the volume of funding, mainly due to the rise of (about 24\%) interconnecting neighbours  (Fig.~\ref{fig:richcore}d). Overall, we also observed a higher level of interconnectedness among the neighbours of Heriot-Watt University. The two different collaboration strategies we
observed in response to changes in funding policy appeared
to be both effective ways in securing research grants as, among the well funded affiliations we examined, both the top 10 and those
less funded, have directly benefited from their respective 
strategy as their awarded sums continued to rise with the available funds
(Fig.~\ref{fig:richcore}e and Supplementary Fig. S4 online).

\subsection*{The rich core of leading universities}

\begin{figure}[!htb]
\centering
\includegraphics[width=14cm]{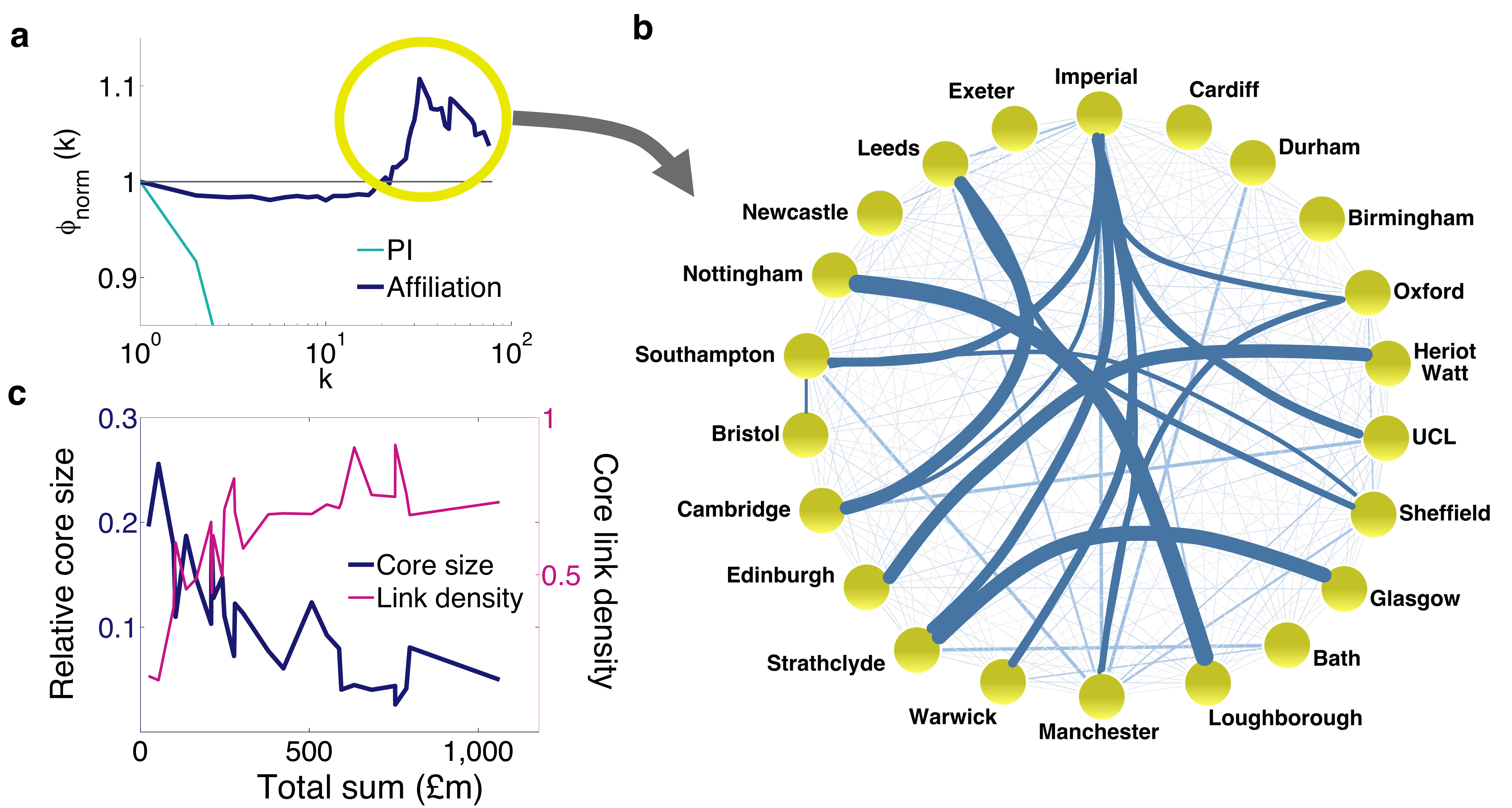}
\caption{\textbf{Elitism in scientific collaborations. }(a). The affiliation network exhibited
  a rich-club phenomenon, with
  $\phi_{norm}(k) > 1$ for high degree nodes, while such a behaviour was
  not present in the PI network (the year 2010 is shown; other years
give the same results). (b). The rich core for year 2010 showing the cohesive community formed by the high degree nodes. Each edge is weighted by the frequency of the partnerships found in that year. (c). The relative size of the rich core of the affiliation
  network, in general, shrank over time while the density of
  partnerships among the affiliations in the rich core continued to
  rise. }
\label{fig:RCncore}
\end{figure}

We examined the level of global interconnectedness among the leading PIs and affiliations, which correspond to the nodes with the largest degree, the so-called hubs, by investigating the presence of a rich-club phenomenon in the networks~\cite{zhou2004rich, colizza2006detecting}. In order to do so, we evaluated the number of edges $E(k)$ among the $N(k)$ nodes of degree larger than a given value $k$, and we computed the rich-club coefficient, $\phi_{norm}(k)$, normalised with respect to the case of random graphs with the same node degrees as in the original networks. A value $\phi_{norm}(k)>1$ for large $k$ is an indication that the hubs of a network have organised into a rich club. Indeed, we observed this phenomenon in the affiliation network, but not in the case of investigators (Fig.~\ref{fig:RCncore}a). This suggests that the affiliations with a high number of grants tended preferentially to collaborate among each others, forming tightly interconnected communities, more than we would expect in a random case.

We applied a core profiling method~\cite{ma2014rich} to extract members of the rich club for each of the 28 years and obtained a total of 45 affiliations (see Supplementary Table~S1 online). Imperial College and University of Oxford were among those in the so-called rich core during the entire period of study, closely followed by other frequently found members which were only absent during the early years and have remained in the rich core ever since. Figure~\ref{fig:RCncore}b shows the rich core in 2010, which was comprised of predominantly the leading affiliations. In particular, the top funded affiliations were not only well connected with the rest of the core, but also showed strong interlinkages among themselves.

We also found that the relative size of the rich core decreased gradually over time (Fig.~\ref{fig:RCncore}c). The rich core initially contained over 25\% of the affiliations and the rise in focussed funding since the start of the millennium has caused the rich core to further shrink and to be maintained to a relatively small size since (4-8\%), and this agrees with existing theory that stress in a system is often manifested in a reduction in core size~\cite{csermely2013structure,liu2011controllability}. The emergence in focussed funding, however, has fostered higher interlinkage among the affiliations in the rich core, as the density of links was initially found to be low and gradually increased during the same period. The density was found to be the highest when the rich core was the smallest. The presence of a rich core of the leading affiliations is a reflection of close collaborations among their researchers, and the increased level of  interlinkage among these affiliations demonstrates a tendency for their researchers to collaborate with their peers in comparably reputable places, suggesting homophily~\cite{mcpherson2001birds,kossinets2009origins} with respect to affiliation excellence accounts for one of the driving forces of scientific collaborations.

\subsection*{Research performance}

\begin{figure}[!htb]
\centering
\includegraphics[width=17cm]{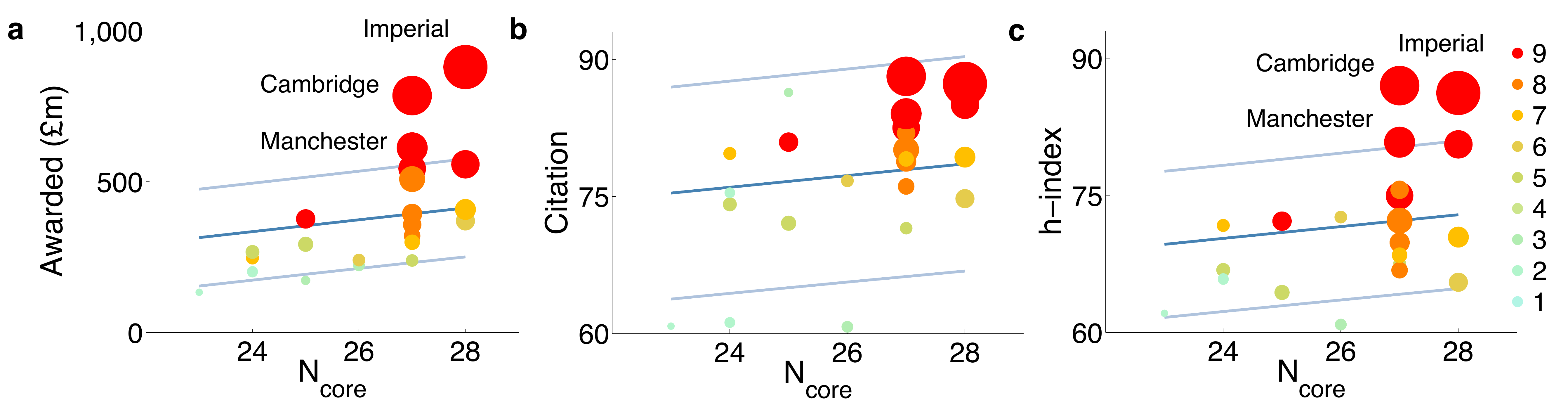}
\caption{\textbf{Fundings and research performance of the rich core.}
(a). The total funding received by each university over the
  entire 28 years is plotted as a function of the number of
  times an affiliation was found in the rich core, $N_{core}$. Node size is proportional to
  the total funding of the affiliation, while the colour denotes the
  number of active research areas which have received a score. Dark
  and light blue lines denote the mean and one standard deviation from
  the mean respectively. (b). The citation scores of the most frequently found affiliations increased with their appearance in the rich core. (c). Similarly, the appearance in the core is indicative of an affiliation's h-index.}
\label{fig:performa}
\end{figure}

Could being in the rich-core a key towards success in research? We investigated in terms of inputs versus outputs of its member affiliations by examining the amount of funding they received and their overall research performance~\cite{jones2008multi,deville2014career}. Firstly, we found that the awarded sum of an affiliation over the 28 years linearly correlated with the number of times, $N_{core}$, the affiliation was present in the rich core (Fig.~\ref{fig:performa}a and Supplementary Fig. S5 online), 
with the exception of Imperial College, University of Cambridge and University of Manchester which have accumulated
funding beyond what is expected from a linear behaviour between $N_{core}$
and total fundings. In addition, we referred to the number of research areas of an affiliation as a measure of the {\em breadth} of its research, and we observed this quantity generally declines as $N_{core}$ decreases.  

We also considered the relative citation score~\cite{citation} and the h-index~\cite{hirsch2005index, hirsch2007does} of each affiliation to capture the {\em quantity} and the {\em depth} of research respectively; the former gauges the volume of citations among researchers in an affiliation, and the latter additionally captures quality. We found that the citation score increased linearly with $N_{core}$ (Fig.~\ref{fig:performa}b). The absence of an obvious deviation from the average trend here suggests the capital requirement on developing a full breadth of
research areas is nontrivial; while the top 3 affiliations appeared to have successfully done so by scoring in all areas, they did not receive citations exceedingly. 
As for the h-index, again we observed that Imperial College, University of Cambridge and University of Manchester outperformed the rest of the universities in this metric, showing a marked deviation from the linear behaviour (Fig.~\ref{fig:performa}c) in all similar to that observed in the awarded sum. Their outstanding funding profiles appeared to have enabled them to develop the depth of research which has led to the generation of high quality papers; and University of Cambridge was particularly the case with the highest average h-index.

\section*{Discussion}

There is today an open question on the kind of collaboration mechanics that underpin success in science, and up to now, such an issue has been mainly explored by characterising the structural patterns of collaborations based on how scientists publish together~\cite{moody2004structure,newman2004best,newman2004coauthorship}. Our approach complements those studies based on coauthorship networks, addressing the question by looking at how scientists collaborate in order to obtain fundings for their research.

Our results demonstrate that there has been a striking change in the funding landscape in the UK in the last three decades, with the largest engineering and physical sciences funding agency providing focused research investment to selected places. This has also shaped the anatomy of collaborations among investigators and among universities in terms of both the local cohesiveness and global interlinkage among them. In particular, the network analysis of successful project partnerships among affiliations has shown that elite universities have been able to capitalise on this competitive and rapidly changing environment; as they have become extensively the brokers of the network, orchestrating partnerships through a diverse source. This is likely to have arisen due to their ability to develop a broad range of expertise and extend partnerships with specialist entities~\cite{jones2009burden}. In addition to that, elite universities formed the very centre of a rich core through their strong reliance. 

Indeed, our findings show the presence of an elite group of affiliations over-attracting resources, and many in the research community would find such inequality in funding distribution highly controversial~\cite{fortin2013big}. However, the effect appeared not to be entirely adverse, as arguably, those elite affiliations which have successfully become very rich seemed to have repaid in both the variety of research and unmistakably in its quality. The prominent role of elite universities is often said to have a wider impact in driving science forward~\cite{jones2008multi,clauset2015systematic}. Here, this is demonstrated by the fact that other well-funded places, which might have less capacity to expand, have consistently benefited from their association with the elites through the rich core. This agrees with previous claims on the importance of elite universities in facilitating multi-partner collaborations~\cite{jones2008multi}. 

Here, we revealed how collaboration networks constantly undergo adaptive organisation in response to funding, resulting in an evident shift in network configuration across different scales. These shifts in collaboration patterns may potentially mean a change in the nature of these partnerships, altering their strengths (e.g. frequency and financial value). This line of research remains unexplored, and hence, a comprehensive study on the weighted version of these networks will provide a leap in understanding the magnitude of these variations in patterns. Furthermore, studying into other attributes that may also covary with funding, such as the degree of interdisciplinary research, will allow us to gain an insight into how mechanisms~\cite{pan2012world,sun2013social} for formulating partnerships have evolved, as the impact of funding is highly likely to be multidimensional.

Funding is an essential part of research as it provides the very much sought after resources to support related activities. Applying for funding is highly competitive~\cite{Cimini2014} as suggested by the low average success rate, and in many countries the economic recession observed in recent years has led to considerable pressure on the research budget. A better understanding on successful project formations, therefore, provides an insight into how to thrive in such challenging landscape.

\section*{Acknowledgements}
AM acknowledge the partial financial support from ImpactQM. We thank Ursula Martin, David Arrowsmith, Pietro Panzarasa and Mario Chavez for their comments and discussions.

\newpage

\section*{Methods}

\subsection*{Dataset - EPSRC Grants on the Web (GoW) }
The EPSRC, formerly the Science and Engineering Research Council (SERC), was created in 1981 to reflect the increased emphasis on engineering research. The dataset was collected from the EPSRC GoW system~\cite{EPSRCGoW} which, at the time of data collection, contained a total of 43,193 projects awarded between 1985 and 2013 (only grants starting prior to 1st Oct 2013 were available). For each project, we recorded the title and the total value, the Principal Investigator (PI), any Co--investigator(s) (CIs) and/or other investigator(s) (OIs), the start and end dates, affiliation(s) of the investigators(s), and the associated research area(s). Funding was considered to be awarded to the PI and the affiliation of the PI. Information on how the overall funding of a given grant was divided among the rest of the investigators (and their affiliations) was not made available. 

\subsection*{Adjustment to monetary values }
As the period of study spans over three decades, the Consumer Prices Index (CPI) with respect to 2005, provided by the Office for National Statistics~\cite{ONS_CPI}, has been applied to adjust the figures so that the monetary values of grants were made comparable across the whole period of study. This index was chosen as the costing associated with a grant is often originated from salaries, travel and equipment (including consumables), and their changes in value over time are well captured by the CPI.

\subsection*{Network construction}
Two types of networks were constructed based on the EPSRC GoW. Firstly, we referred to a project partnership between a PI and a collaborator (a CI or an OI) as an edge and the resultant network was referred to as the investigator network. Grants consisted of only one investigator, i.e. only the PI, have been excluded from the network construction. Similarly, we referred to a project partnership between the affiliations of a PI and a collaborator (a CI or an OI) as an edge and obtained the affiliation network. In both cases, we constructed a network for each year by referring to all the active projects and in this case an edge would be present for the duration of a project. For example, if a project started and ended in 1990 and 1993, the associated edges would be present between 1990 and 1993 inclusively. All networks were studied as undirected and unweighted graphs.

\subsection*{Effective size $\zeta_{i}$ }
To examine the degree of brokerage of an affiliation we focused on the presence of structural holes~\cite{ronald1992structural} in the network: a node acts as an information broker when it bridges otherwise disjoint neighbours. We assessed the level of brokerage of a node $i$ by calculating its normalised effective size as: $\zeta_{i} = 1-\frac{k_i - 1}{k_i} C_i$ ~\cite{ronald1992structural,latora2013social}, where $k_i$ is the degree of the node and $C_i$ is its clustering coefficient. $\zeta_{i}$ ranges between 0 and 1. A  value of 0 denotes a fully connected local network, implying a fully cohesive clique; while a value of 1 reflects a centralised local network with a star configuration which means the highest level of brokerage.

\subsection*{The rich-club phenomenon and the rich core }
To detect the presence of a rich-club phenomenon we first ranked the network nodes in descending order of their degree, $k$.
Starting with the group of nodes with the highest
degree, we calculated for each value of $k$ the
quantity $\phi(k) = \frac{2E(k)}{k(k-1)}$,  
where $E(k)$ is the number of edges among nodes
with a degree $\geq k$~\cite{zhou2004rich}.
To determine $\phi_{norm}(k)$, we compared $\phi(k)$ of the empirical network to that obtained from an ensemble
of random graphs. Specifically, for a given empirical network, we constructed
1000 random graphs with the same degree sequence as the empirical
network and obtained
${\phi_{rand}(k)}$ by averaging over the ensemble of networks. Finally,
$\phi_{norm}(k) = \frac{\phi(k)}{\phi_{rand}(k)}$~\cite{colizza2006detecting}.

Once we have established that high degree affiliations exhibited a rich-club phenomenon, we characterised the membership of the club in each year by applying a core profiling method~\cite{ma2014rich}. We ranked the importance of  the nodes in descending order of their degree, such that the node with the highest degree is ranked first and so on. For a given node, we divided its links into two groups: those with nodes of a higher rank and those with a lower rank. More formally, a node with a rank $r$ has degree $\degree_r$; the number of links it shares with nodes of a higher rank is $\linksRC_r$ and the number of links with nodes of a lower rank is $\degree_r-\linksRC_r$. Starting from the node with the highest rank, as $r$ increases the number of links $\linksRC_r$ that node $r$ shares with nodes of a higher rank fluctuates. There will be a node $r^*$ where $\linksRC_r$ has reached its maximum, and from that node onwards $\linksRC_r$ is always less than $\linksRC_{r^*}$. This change in the connectivity among the highly ranked nodes defines the boundary of the rich core: the nodes with a rank less than or equal to $r^*$ are the core and the rest belong to the periphery~\cite{ma2014rich}.

\subsection*{Research performance}
We evaluated the research performance of an affiliation by its relative citation score~\cite{citation} and the h-index~\cite{hirsch2005index, hirsch2007does} in 2013. The former accounts for the total number of citations per faculty member in an affiliation, and is  considered as a measure of the volume of research (i.e. quantity). The latter takes into consideration the citations of the best papers, and is considered as a measure of the depth of research (i.e. quality). The scores for individual affiliations were obtained from the QS World University Rankings\cite{QSranking} and they were calculated based on bibliometric data on Scopus~\cite{scopus}. Both quantities were normalised to take values in the range $[0,100]$. 
Scores were available in a range of disciplines and only subject areas related to the Engineering and Physical Sciences domains were selected, including Chemistry, Computer Science and Information Systems, Engineering (Chemical, Civil and Structural, Electrical and Electronic, and Mechanical), Materials Sciences, Mathematics and Physics. Individual subject citation scores and h-indexes of a given affiliation were averaged and used as a measure of performance. Only Imperial, Cambridge, Manchester, Oxford, UCL and Edinburgh have scored in all areas. 

\newpage
\beginsupplement
\section*{Supplementary Materials}
\subsection*{Figures}

\begin{figure}[!htb]
\centering
\includegraphics[width=16cm]{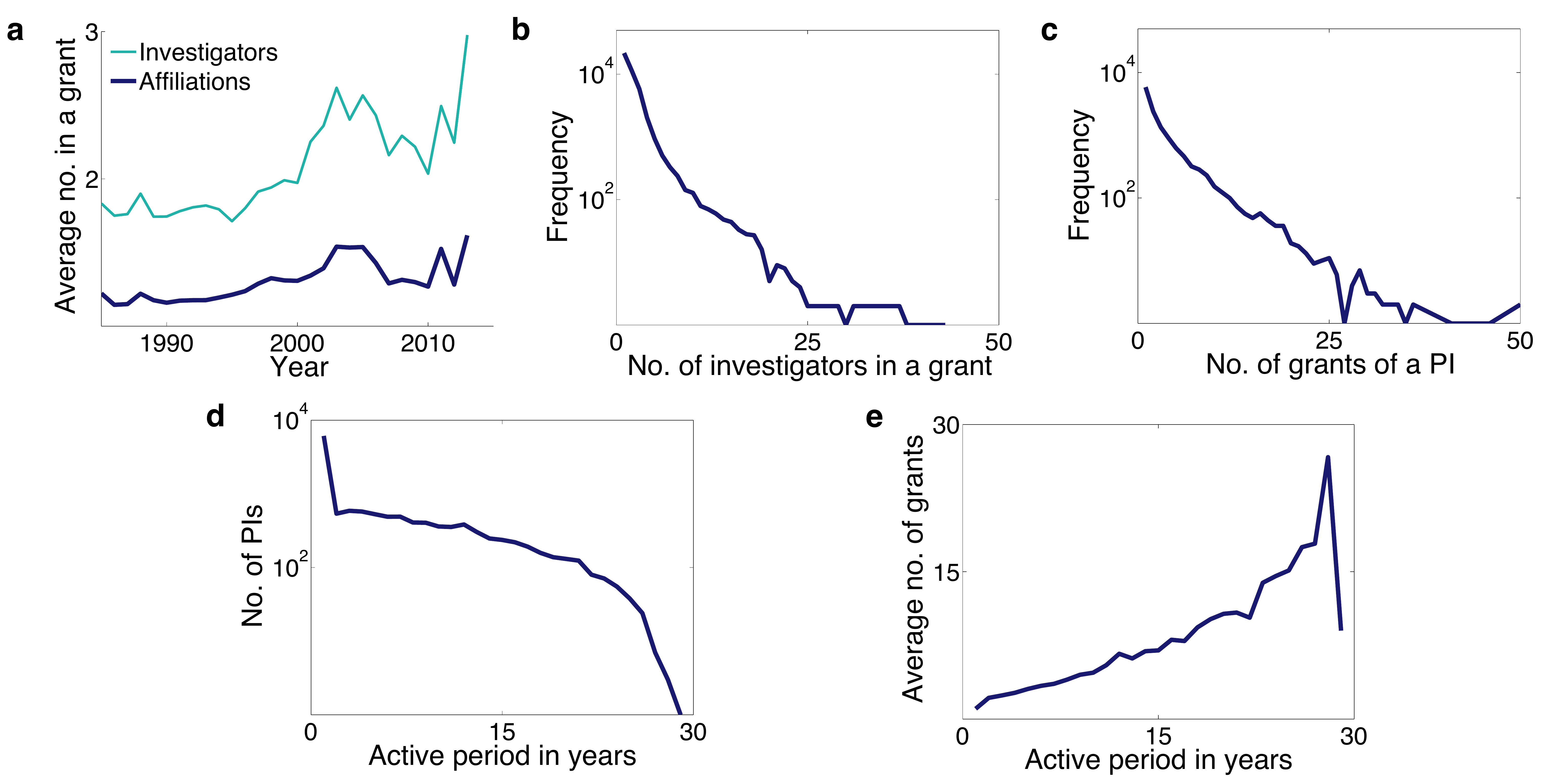}
\caption{\textbf{Patterns between grants and principle investigators. }(a). The average number of investigators and associated affiliations generally increased over time. (b) The number of investigators in a grant (include all PIs, CIs and OIs) and its frequency. (c) Most PIs had only one successful grant and the number of grants a PI received declined sharply; for example, there were less than 10 PIs who have managed to obtain more than 20 grants. (d) The number of PIs against the number of active years which is defined as the sojourn time between the first and last recorded grants of a given PI. (e) The average number of grants obtained by a PI generally increased with his/her career length.}
\label{fig:investigator}
\end{figure}
About half of the grants (21954 to be precise) were associated with one or more CIs and/or OIs. Both the average numbers of investigators and affiliations increased over time (Fig.~\ref{fig:investigator}a). While a grant could have over 40 investigators, by large, the majority of grants consisted of a small number of investigators (Fig.~\ref{fig:investigator}b). The number of grants awarded to a PI decreased exponentially (Fig.~\ref{fig:investigator}c). We defined an active period of a PI as the number of years between his/her first and last recorded grants. Fig.~\ref{fig:investigator}d shows a gentle decline on the number of PIs as the number of active year increased. We observed a direct correction between the length of active research period and the number of grants one obtained (Fig.~\ref{fig:investigator}e).

\newpage

\begin{figure}[!htb]
\centering
\includegraphics[width=16cm]{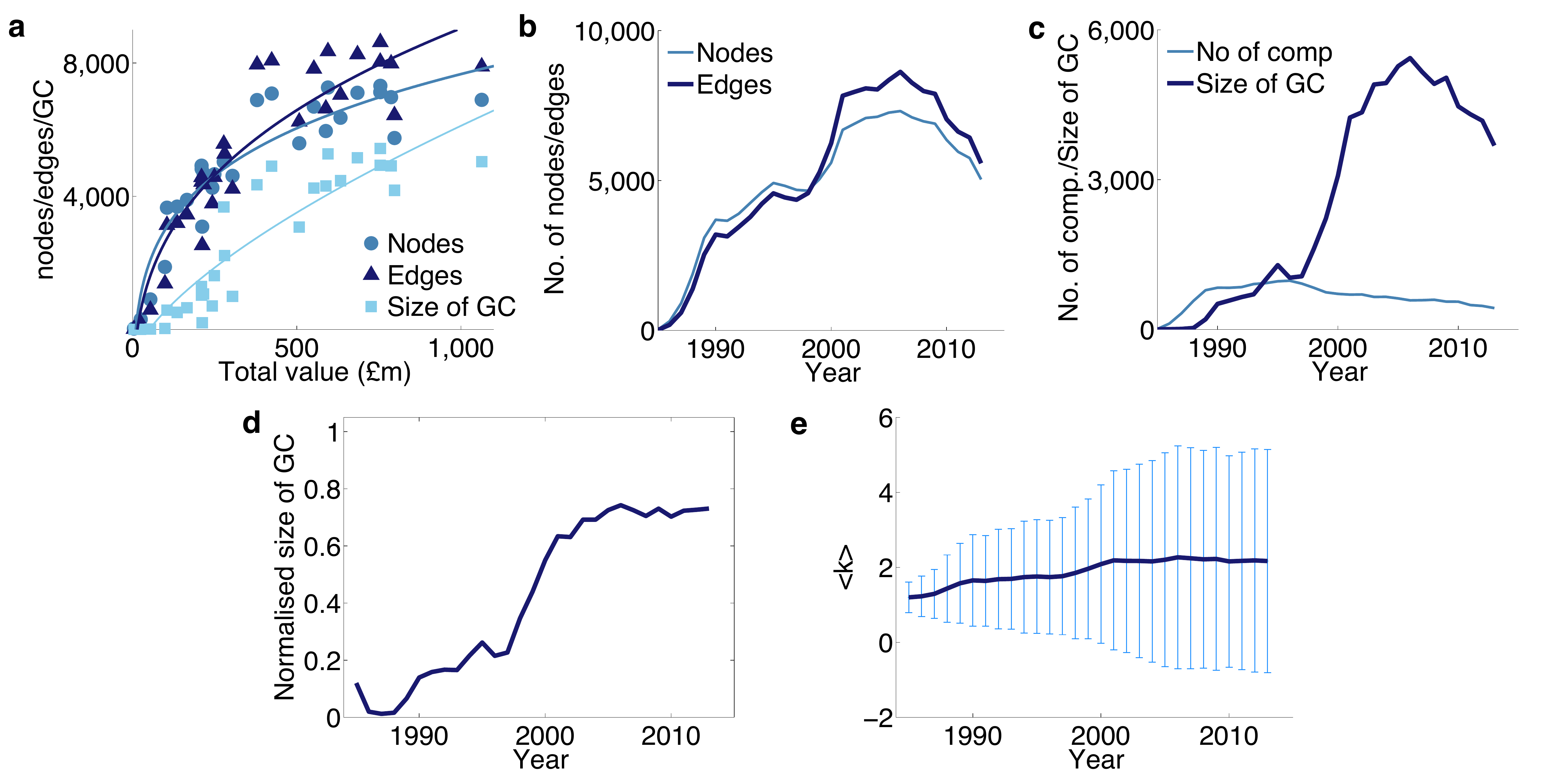}
\caption{\textbf{Network properties of the investigator network. }(a). The number of nodes and edges and the size of the giant component (GC) covaried with the total funding awarded by the EPSRC of a given year. (b). The number of nodes and edges in the investigator network over time. (c). The number of components in the network and the size of the giant component. (d). The giant component grew over time; the normalised size of the GC is obtained by dividing the size of the giant component  by the total number of nodes. (e). The average degree remained very low throughout the period of study. }
\label{fig:PI_CIOI_org}
\end{figure}
In the network of investigators, the network size in terms of both number of investigators and collaborations correlated with funding (Fig.~\ref{fig:PI_CIOI_org}a). Though, we observed a threshold at about \pounds500 million in which network growth slowed down with further increase in funding. The total funding reached over this threshold in 2000 (Fig.~\ref{fig:PI_CIOI_org}b); and in fact, the network reduced in size when there was a noticeable rise in funding in 2009. Similarly, an increased funding also resulted in a larger giant component in the network, implying better interconnectivity among different local groups of investigators (Fig.~\ref{fig:PI_CIOI_org}c and d); though the overall connectivity of the network remained sparse as the average degree was low and variable (Fig.~\ref{fig:PI_CIOI_org}e).

\newpage
\begin{figure}[!htb]
\centering
\includegraphics[width=16cm]{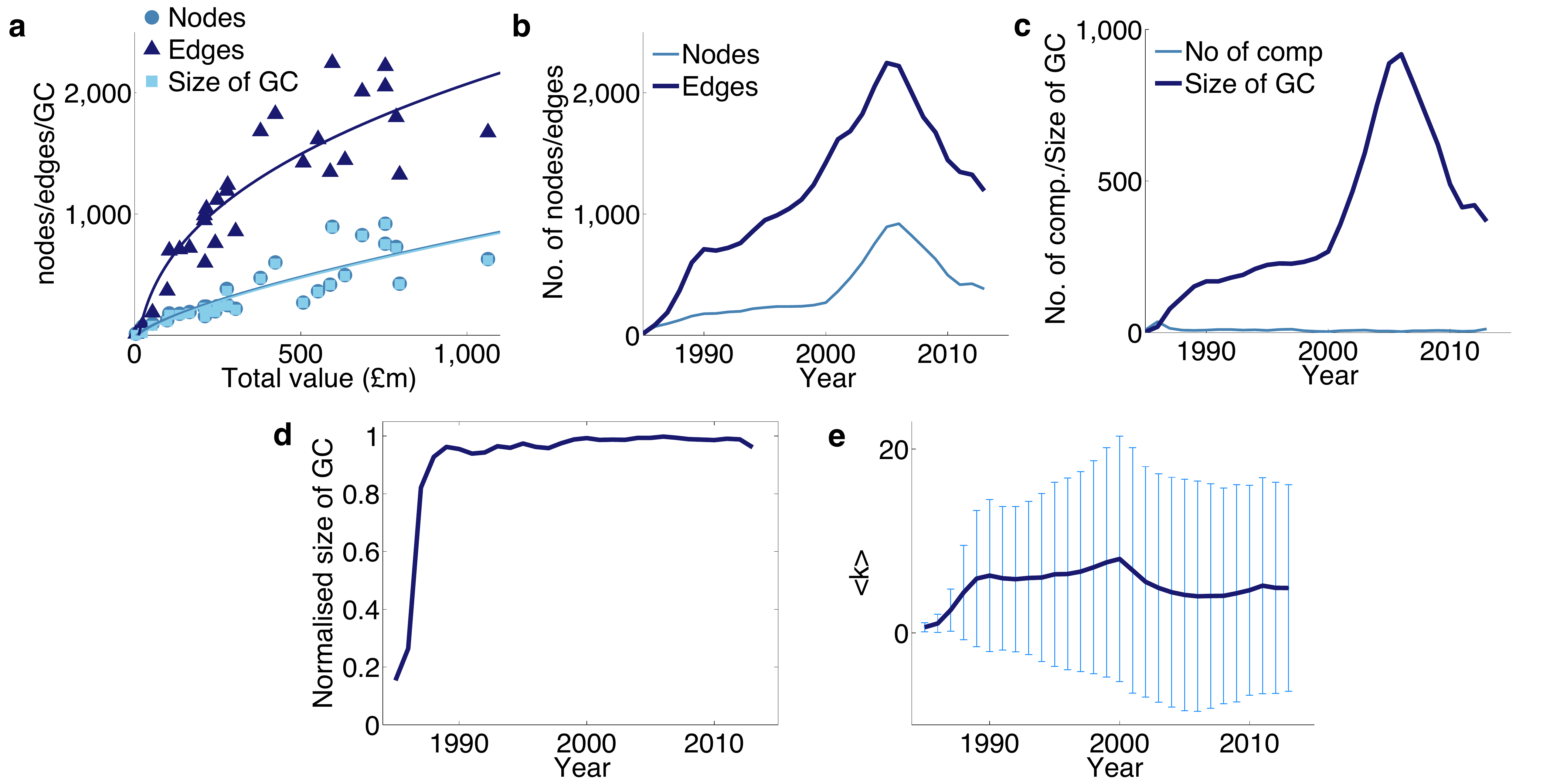}
\caption{\textbf{Network properties of the affiliation network. }(a). The number of nodes and edges and the size of the giant component (GC) covaried with the total funding. (b). The number of nodes and edges in the affiliation network over time. (c). The number of components in the network and the size of the giant component. (d). The network has remained well connected as the size of the giant component remained equal to or very close to the size of the network. (e). The average degree was variable.  }
\label{fig:PI_CIOI_orgB}
\end{figure}

We observed similar covariance between funding and network size (Fig.~\ref{fig:PI_CIOI_orgB}a) in the network of affiliations. Again, the network shrank in size as a result of the sharp increase in funding in 2009 (Fig.~\ref{fig:PI_CIOI_orgB}b); this provides another evidence on heterogeneous distribution of funding. In contrast to the investigator network, the affiliation network was well connected with very few disjoint components (Fig.~\ref{fig:PI_CIOI_orgB}c and d). The average degree was significantly higher than that of the investigator network, and of much higher level of variability (Fig.~\ref{fig:PI_CIOI_orgB}e).

\begin{figure}[!htb]
\centering
\includegraphics[width=16cm]{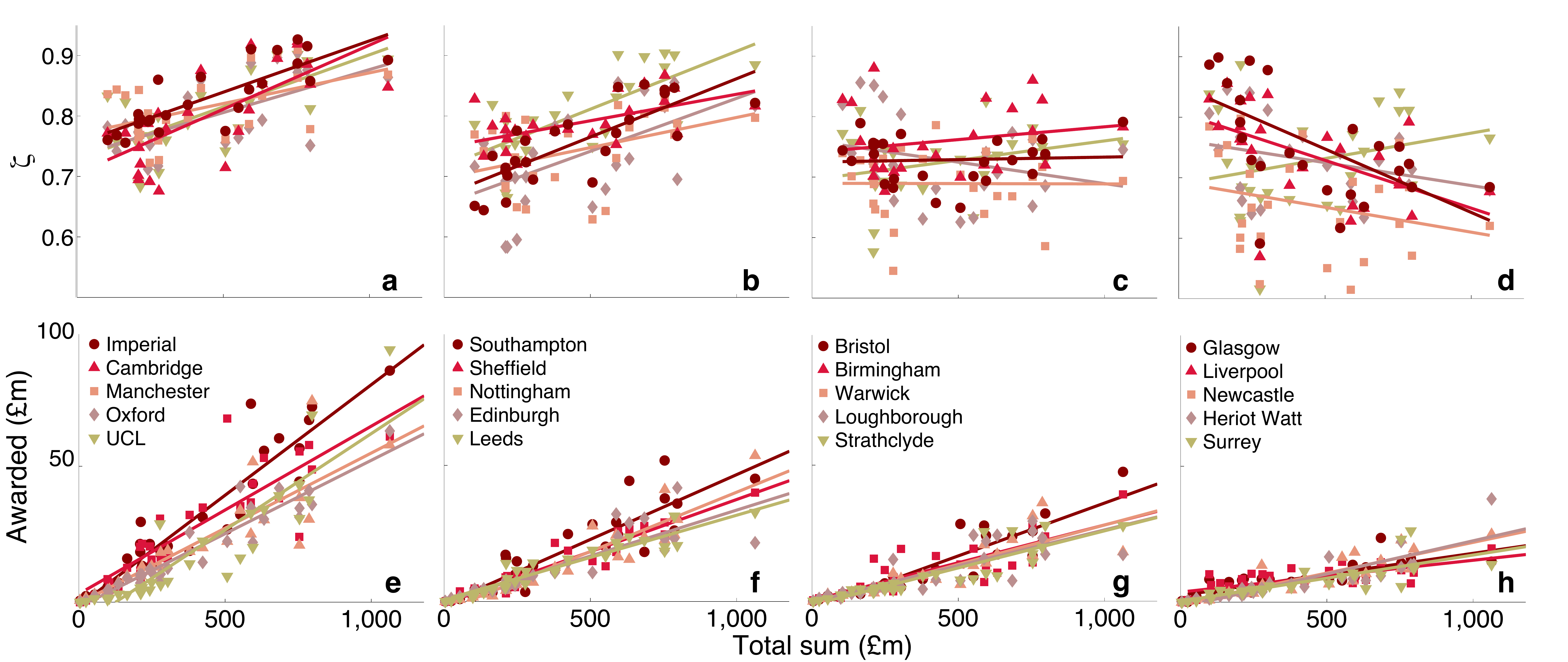}
\caption{\textbf{Collaboration patterns of the top 20 funded affiliations covaried with funding.} This figure shows the effective size and awarded funding of the top 20 funded affiliations which is an extension of Fig. 2 (panel (b) and (e)) in the main paper, in which only results of Imperial and Heriot-Watt are shown. (a-b) Brokerage among the top 10 affiliations clearly increased with the total funding. Other less well funded affiliations showed a less distinct trend (c) while most affiliations in the last quantile decreased in brokerage as the funding increased (d).
(e-h) Funding obtained by the top 20 affiliations continued to increase with the total funding. }
\label{fig:Orgfundnet}
\end{figure}

\begin{figure}[!htb]
\centering
\includegraphics[width=16cm]{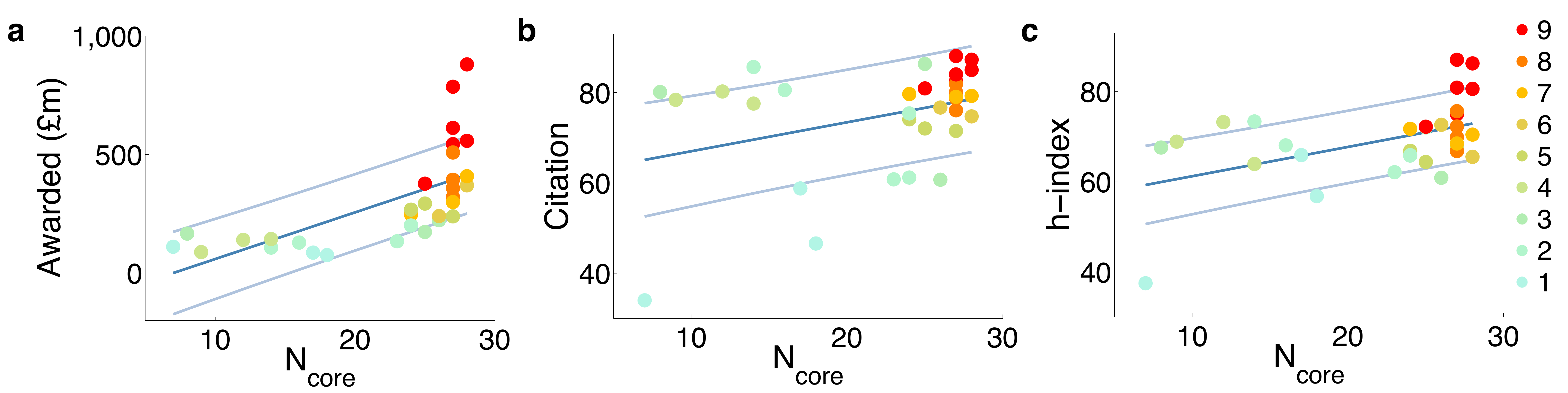}
\caption{\textbf{Funding and research performance of the rich core.} This figure is the full version of Fig. 4 in the main paper, which only shows affiliations with $N_{core}>20$. (a). The more an affiliation appeared in the core, the more funding it was likely to received. Colour denotes the number of active research areas which have received a score. Dark and light blue lines denote the mean and one standard deviation from the mean respectively. (b). Most frequently found affiliations in the rich core; and their citation score increased with their appearance in the rich core. (c). Similarly, the appearance in the core is indicative of an affiliation's h-index.}
\label{fig:coreperSI}
\end{figure}

\clearpage
\subsection*{Tables}

\begin{longtable}{ | p{5.5cm} | p{2.5cm} | r |r | r | r |}
    \hline
    \bf{Affiliation} & \bf{Abbreviation} & \bf{Grants} & \bf{Total \pounds} & \bf{$N_{core}$} & \bf {$\zeta_{avg}$}  \\ \hline    
     \endhead
Imperial College	 & 	Imperial	 & 	2658	 & 	880,042,691	 & 	28	 & 	0.847	 	\\ \hline
University of Cambridge	 & 	Cambridge	 & 	2182	 & 	785,600,273	 & 	27	 & 	0.806	 	\\ \hline
University of Manchester	 & 	Manchester	 & 	2351	 & 	612,256,769	 & 	27	 & 	0.830		\\ \hline
University of Oxford	 & 	Oxford	 & 	1966	 & 	557,758,967	 & 	28	 & 	0.818	 	\\ \hline
University College London	 & 	UCL	 & 	1417	 & 	543,449,634	 & 	27	 & 	0.828	 	\\ \hline
University of Southampton	 & 	Southampton	 & 	1373	 & 	508,545,865	 & 	27	 & 	0.763	 	\\ \hline
University of Sheffield	 & 	Sheffield	 & 	1367	 & 	408,240,769	 & 	28	 & 	0.803	\\ \hline
University of Nottingham	 & 	Nottingham	 & 	1411	 & 	393,624,426	 & 	27	 & 	0.755		\\ \hline
University of Edinburgh	 & 	Edinburgh	 & 	1191	 & 	376,515,957	 & 	25	 & 	0.747		\\ \hline
University of Leeds	 & 	Leeds	 & 	1452	 & 	369,915,734	 & 	28	 & 	0.822	 	\\ \hline
University of Bristol	 & 	Bristol	 & 	1259	 & 	357,784,563	 & 	27	 & 	0.751		\\ \hline
University of Birmingham	 & 	Birmingham	 & 	1059	 & 	320,365,372	 & 	27	 & 	0.784	 	\\ \hline
University of Warwick	 & 	Warwick	 & 	1111	 & 	299,106,838	 & 	27	 & 	0.712	 	\\ \hline
Loughborough University	 & 	Loughborough	 & 	971	 & 	292,614,129	 & 	25	 & 	0.747		\\ \hline
University of Strathclyde	 & 	Strathclyde	 & 	1004	 & 	267,072,205	 & 	24	 & 	0.746	 	\\ \hline
University of Glasgow	 & 	Glasgow	 & 	857	 & 	247,098,481	 & 	24	 & 	0.777	 	\\ \hline
University of Liverpool	 & 	Liverpool	 & 	966	 & 	240,086,056	 & 	26	 & 	0.760		\\ \hline
University of Bath	 & 	Bath	 & 	992	 & 	238,851,816	 & 	27	 & 	0.675	 	\\ \hline
Newcastle University	 & 	Newcastle	 & 	813	 & 	222,668,660	 & 	26	 & 	0.748	 	\\ \hline
Heriot-Watt University 	 & 	Heriot Watt	 & 	857	 & 	202,088,932	 & 	24	 & 	0.754	 	\\ \hline
University of Surrey	 & 	Surrey	 & 	783	 & 	200,322,051	 & 	24	 & 	0.726	 	\\ \hline
Durham University	 & 	Durham	 & 	776	 & 	172,691,133	 & 	25	 & 	0.684	 	\\ \hline
University of St Andrews	 & 	St Andrews	 & 	515	 & 	166,131,172	 & 	8	 & 	0.687		\\ \hline
Cranfield University	 & 	Cranfield	 & 	578	 & 	164,867,131	 & 	18	 & 	0.714	 	\\ \hline
Queen's University Belfast	 & 	Queen's	 & 	660	 & 	142,904,945	 & 	14	 & 	0.712	 	\\ \hline
Queen Mary, Uni. of London	 & 	QMUL	 & 	618	 & 	139,471,209	 & 	12	 & 	0.685	 	\\ \hline
Cardiff University	 & 	Cardiff	 & 	648	 & 	133,551,032	 & 	23	 & 	0.740	 	\\ \hline
University of York	 & 	York	 & 	611	 & 	128,477,800	 & 	16	 & 	0.703		\\ \hline
Swansea University	 & 	Swansea	 & 	488	 & 	110,663,780	 & 	7	 & 	0.735		\\ \hline
Lancaster University	 & 	Lancaster	 & 	516	 & 	107,044,160	 & 	14	 & 	0.682	 	\\ \hline
Brunel University	 & 	Brunel	 & 	415	 & 	91,665,418	 & 	18	 & 	0.714		\\ \hline
King's College London	 & 	King's	 & 	530	 & 	88,242,899	 & 	9	 & 	0.708	 	\\ \hline
University of Exeter	 & 	Exeter	 & 	454	 & 	86,135,975	 & 	17	 & 	0.618	 	\\ \hline
University of Reading	 & 	Reading	 & 	467	 & 	78,721,875	 & 	17	 & 	0.682		\\ \hline
University of Salford	 & 	Salford	 & 	470	 & 	75,592,400	 & 	18	 & 	0.770		\\ \hline
STFC - Laboratories	 & 	STFC	 & 	215	 & 	69,769,793	 & 	12	 & 	0.589	 \\ \hline
University of Leicester	 & 	Leicester	 & 	382	 & 	48,127,442	 & 	1	 & 	0.585	 	\\ \hline
University of Kent	 & 	Kent	 & 	335	 & 	45,236,232	 & 	1	 & 	0.678	 	\\ \hline
University of East Anglia	 & 	E. Anglia	 & 	308	 & 	42,871,012	 & 	2	 & 	0.652		\\ \hline
University of Hull	 & 	Hull	 & 	320	 & 	41,882,084	 & 	1	 & 	0.655	 	\\ \hline
University of Dundee	 & 	Dundee	 & 	236	 & 	40,908,598	 & 	5	 & 	0.690		\\ \hline
University of Bradford	 & 	Bradford	 & 	287	 & 	36,219,345	 & 	1	 & 	0.737	 	\\ \hline
Royal Holloway, Uni. of London	 & 	R. Holloway	 & 	191	 & 	34,035,831	 & 	1	 & 	0.675	 	\\ \hline
Keele University	 & 	Keele	 & 	213	 & 	22,234,981	 & 	1	 & 	0.670	 	\\ \hline
Bangor University	 & 	Bangor	 & 	193	 & 	18,407,971	 & 	1	 & 	0.582	 	\\ \hline
\caption{\textbf{Members of the rich core.} Affiliations were sorted in descending order of their total funding. For each affiliation, their total number of grants (Grants) and total amount funding awarded (Total $\pounds$), frequency found in the rich core ($N_{core}$) and yearly average effective size ($\zeta_{avg}$) are shown. 1985 has been omitted as the network only had 8 affiliations.}
\label{table:top37}
\end{longtable}

\end{document}